\documentstyle[12pt,psfig]{article}
\def \LSP{\widetilde{\chi_1}^0}
\def \N2{\widetilde{\chi_2}^0}
\def \CH{\widetilde{\chi}^{\pm}}
\def \W1{\widetilde{\chi}_1^{\pm}}   
\def \WC2{\widetilde{\chi}_2^{\pm}}

\def \SELL{\tilde{e}_L}

\def \CH{\widetilde{\chi}^{\pm}}
\def \W1{\widetilde{\chi}_1^{\pm}}
\def \WC2{\widetilde{\chi}_2^{\pm}}

\def \SELL{\tilde{e}_L}

\def \DR{\tilde{d}_R}

\def \MN2{m_{\widetilde{\chi_2}^0}}
\def \MCH{m_{\widetilde{\chi}^{\pm}}}
\def \MCH1{M_{\widetilde{\chi}_1^{\pm}}}

\def \MDR{m_{\tilde{d}_R}}
\def \MGLU{m_{\tilde{g}}}
\def \MSQ{m_{\tilde{q}}}
\def \MER{m_{\tilde{e}_R}}
\def \MEL{m_{\tilde{e}_L}}
\def \MUL{m_{\tilde{u}_L}}
\def \MUR{m_{\tilde{u}_R}}
\def \MDR{m_{\tilde{d}_R}}
\def \ET{\not\!\!{E_T}}

\def \leqv{\stackrel{<}\sim}
\begin{document}
\setcounter{page}{0}
\thispagestyle{empty}
\begin{flushright}
MRI-P-000602\\
hep-ph/yymmxxx
\end{flushright}
\begin{center}
{\large\bf Signatures of the Light $\DR$ Scenario at the Upgraded 
Tevatron}\\
\bigskip
{\normalsize {\sf Amitava Datta}{\footnote {adatta@juphys.ernet.in}}}\\
{\footnotesize 
Department of Physics, Jadavpur University, Calcutta 700 032,
India.}\\
{\normalsize {\sf Shyamapada Maity} {\footnote {shyama@juphys.ernet.in}}}\\
{\footnotesize 
Tamralipta Mahavidyalaya, Tamluk 721636, West Bengal, India.}\\
{\normalsize {\sf Anindya Datta} {\footnote {anindya@mri.ernet.in}}}\\
{\footnotesize 
Mehta Research Institute, Allahabad 211019, India.}\\
\end{center}
\vskip 5pt
\begin{center}
{\bf ABSTRACT }
\end{center}

In scenarios with relatively light down squarks, motivated, e.g., by
$SO(10)$ D terms, jets + $\ET$ signals can be observed at the luminosity
upgraded Tevatron even if the squarks are much heavier than the
gluinos and the common gaugino mass ($M_{\frac{1}{2}}$) at $M_G$
lies above the LEP allowed lower bound . In the conventional mSUGRA
model with heavy squarks practically no signal is expected in this
channel. The possiblity of distinguishing between various SUGRA
motivated scenarios by exploiting the $\ET$ and jet $p_T$
distributions, opposite sign dileptons + jets + $\ET$ events and clean
trilepton signals have been discussed. 

\vskip 10pt
\section{Introduction}

The search for supersymmetry (SUSY) \cite{1} has been going on at the
leading high energy colliders, most notably at LEP and
Tevatron (Run-I), for quite some time.  From the negative results lower
limits on various sparticle masses have been obtained \cite{2}. The
prospect of SUSY searches at Tevatron(Run-II) and at the large hadron
collider (LHC) has also been studied in great details \cite{3,4}. The
sparticle mass reach of these colliders in different channels have
also been estimated.

In most of the analyses it is assumed, for the sake of economy in the
number of parameters, that all the scalars in the model, i.e. the
squarks, the sleptons and the Higgs bosons, have a common SUSY
breaking mass ($m_0$) at the grand unified theory (GUT) scale
($M_G$). Moreover the gaugino masses and the trilinear soft breaking
terms are also assigned common values $m_{\frac{1}{2}}$ and $A_0$
respectively, at $M_G$.  The parameters at the energy scale of
interest ($\sim $ few hundred GeV) is determined by the usual
renormalisation group (RG) running.

The number of free parameters may be further reduced by requiring
radiative $SU(2)\times U(1)$ breaking at the electroweak scale. This
fixes the magnitude of the Higgsino mass parameter ($\mu$). Thus
$m_0$, $m_{\frac{1}{2}}$, $A_0$ along with the sign of $\mu$ and $\tan
\beta$ (the ratio of the vacuum expectation values of the two neutral
higgs bosons) define the model completely. This popular model is
hereafter referred to as the conventional scenario.

The above framework motivated by N=1 supergravity \cite{5} is very
attractive. However as there is no direct experimental information
about physics at $M_G$, it is imprudent to restrict our attention to
this model only.  In this paper our goal is to re-examine the
prospective SUSY signals at the Tevatron (Run-II) by relaxing some of
the above assumptions. We shall, however, assume that the gaugino
masses unify at $M_G$. This assumption is quite natural within the
framework of any SUSY GUT, since it follows if the GUT symmetry is
respected by the SUSY breaking mechanism at some high scale.

The assumption of a common soft breaking mass $m_0$ at $M_G$ is
undoubtably more model dependent. Unlike the gauginos different
scalars in a SUSY GUT may belong to different representations of the
GUT group. This is especially so for the light higgs scalars and the
sfermions, which almost always reside in different multiplets.  Even
if we assume the validity of the supergravity model, the universal
parameter $m_0$ may well be generated at a scale substantially
different from $M_G$, say the Plank Scale ($M_P$). Then the running of
the scalar masses, belonging to different multiplets of the GUT group,
between $M_P$ and $M_G$ may lead to non-universality at $M_G$
\cite{6}.

The following nonuniversal scenario is rather interesting from the
phenomenological point of view. In this scenario the `right - handed'
down - type squarks ($\tilde{d}_R, \tilde{s}_R, \tilde{b}_R$ ),
generically denoted by $\tilde{d}_R $, are significantly lighter than
the other squarks.  Then the gluino decays into three body final
states mediated by virtual $\tilde{d}_R$ squarks will
dominate. Further if the LSP is assumed to be dominated by the $U(1)$
gaugino ($\tilde{B}$), then practically all of these virtual
$\tilde{d}_R$'s will decay into the LSP and a d-type quark. Thus the
branching ratio ( BR) of direct gluino decays into the LSP will be
enhanced, while the cascade decays of the gluino will be
correspondingly suppressed.  In the special case $\MGLU > \MDR$, while
all other squarks are heavier than the gluino, practically all the
gluinos will decay into the jets + $\ET$ channel with a remarkably
hard $\ET$ spectrum. On the other hand gluino decays into leptons +
jets + $\ET$ arising through cascade decays will be strongly
suppressed. The signal from $\tilde{g}\tilde{g}$, $\tilde{g}\DR$ and
$\DR\DR$ production is likely to be observable, although the other
squarks may be heavy to be of any consequence at Tevatron energies
\cite{7,8}.

Theoretically relatively light $\tilde{d}_R$'s can be naturally
motivated within a SUSY GUT framework in a variety of ways. If the GUT
group is $SU(5)$, then the $\DR$ squarks residing in the 5 - plet may
be renormalized between $M_P$ and $M_G$ such that the resulting soft
breaking mass at $M_G$ is significantly smaller than that of the other
squarks belonging to the 10 - plet \cite{6}. The numerical results of
ref. \cite{6}, though in the right direction, does not exhibit a large
enough mass split.

In this paper we shall illustrate the signatures of a light $\DR$
scenario through an $SO(10)$ SUSY GUT to be discussed below. 
Such models are now much more popular than the good old $SU(5)$
SUSY GUT in view of the recent excitement about neutrino masses and
mixings generated by the SUPERK and other experiments \cite{9}.

We, however, emphasize that the novel collider signatures are
essentially consequences of the above squark - gluino mass hierarchy
at low energies and are fairly insensetive to the details of GUT scale
or Planck scale physics responsible for generating it.  Moreover, in
view of the large uncertainties involved in GUT scale - Plank scale
physics the quantitative results need not be regarded as firm
predictions. Therefore, keeping in mind that either of the above
mechanisms or their combination can in principle generate the required
mass hierarchy, one might as well discuss the resulting phenomenology
in a model independent way.

 We shall now focus our attention on an SO(10) SUSY GUT \cite{10}
 containing all the quarks and leptons of a given generation in a 16
 dimensional multiplet which includes the heavy right handed neutrino.
 In this model the non-universility at $M_G$ due to running between
 $M_P$ and $M_G$, is expected to be negligible for the first two
 generations of squarks and sleptons with small Yukawa couplings.  In
 principle nonuniversal masses for the third generation sfermions with
 a large Yukawa coupling is also possible due to this
 mechanism. However we shall assume this intergeneration
 nonuniversility to be small compared to the nonuniversality due to D
 terms, which will be described below.

Running of the soft breaking masses between $M_P$ and $M_G$ may result
in soft breaking masses of light higgs bosons at $M_G$ significantly
different from that in the sfermion sector. The light higgs doublets
reside in a 10 dimensional representation of SO(10) and hence are
renormalised differently. Moreover they have to couple to other super
heavy GUT fields in order to implement the mass-split between the
coloured higgs bosons and the colour neutral ones responsible for
$SU(2)\times U(1)$ breaking. Unfortunately the magnitude of the
resulting nonuniversility is not calculable without specifying all the
couplings of the higgs bosons, which are not known presently. We,
therefore, do not attempt to study directly the impact of
nonuniversality on higgs phenomenology in this paper. Instead we shall
restrict ourselves to the signature of the squark-glunio production
and decays which are only weakly dependent on the characteristics of
the higgs sector. However, the effect of nonuniversality in the higgs
sector will be taken into account indirectly by treating $\mu$ as a
free parameter.

In summmary we shall work with a SO(10) scenario in which the soft
breaking masses of the squarks and sleptons are equal (= $m_0$) at
$M_G$.  Non-universality at this scale may still arise due to D-term
contributions to the above masses which appear when SO(10) breaks into
a group of smaller rank\cite{11,12}. In general such contributions
could be different for different members of the 16-plet . However,
these non-universal terms are generation independent, so that no
additional problem due to flavour changing neutral currents arise.

As a specific example we shall consider the breaking of SO(10)
directly to the SM gauge group \cite{12}. The group $SO(10)$ contains
$SU(5)\times U(1)$ as a subgroup.  It is further assumed that the
D-terms are linked to the breaking of this $U(1)$ only.  The squark-
slepton masses in this case are\\
\begin{equation} 
\MUL^2 = \MUR^2 = \MER^2 = {m_0}^2 + 0.5D m_0^2 
\end{equation}
\begin{equation} 
\MDR^2 = \MEL^2 = {m_0}^2 - 1.5D m_0^2, 
\end{equation} 

where the unknown parameter D can be of either sign. The mass
differences arise because of the differences in the $U(1)$ quantum
numbers of the sparticles concerned.  As can be readily seen from the
above formula for D $>$ 0, the left handed sleptons ($\SELL$) and right
handed down type squark ($\DR$) are relatively light.  In this paper we
want to concentrate on the collider signatures of the light $\DR$. In
principle the D term contributions to the light higgs masses lead to
further mass splitting between the higgs bosons and the sfermions at
$M_G$. This provides additional motivation for treating $\mu$ as a
free parameter.

The phenomenology of the lighter $\DR$ have been studied by several
authors \cite{7,8}. In this work we shall extend and complement these
studies in several ways.  In \cite{8}, the production cross-section of
squark-glunio pairs and their decay branching ratios were studied.
The effects of the kinematical cuts on the resulting SUSY signals,
however, were not taken into account. In this paper we study the jets
+ $\ET$ as well as opposite sign dileptons + jets + $\ET$ signals by
using a parton level Monte Carlo. We use as a guide line the
kinematical cuts given in \cite{3}, but our main conclusions are
essentially consequences of the spectrum under study and are fairly
independent of the precise choices of these cuts.

Moreover, we shall comment on the sensitivity of the signal on $\mu$
and $\tan \beta$. this important point was not addressed in the earlier
works.  Event generators requiring large amount of computer time,
though essential for precise quantitative studies, are rather
expensive as tools for studying the dependence of the signal on a
large number of parameters. A parton level Monte Carlo on the other
hand enables us to carry out a qualitative study relatively easily.
		
We shall concentrate on two main issues:\\ a) What are the mass
reaches of the upgraded Tevatron in the nonuniversal scenario and how
do they compare with that in the conventional scenario ? \\ b) If a
signal is seen at the upgraded Tevatron, can one distinguish between
the models with lighter $\DR$ and the conventional scenario ?

The plan of the paper is as follows. In section II we shall discuss
regions of the parameter space which are motivated by various
theoretical considerations and are interesting for SUSY searches at
the Tevatron.  In section III we present our result for the jet +
$\ET$ signal. In section IV the discrimination of different models
using the jet + dilepton + $\ET$ and the clean trilepton signal is
presented. Finally in section V the conclusions are summarised.

\section{ The Choice of Parameters and the Overall Strategy  }

As has been stated in the introduction the model under study has the
following  parameters: $ m_0, m_{1/2}, A_0 , \mu$,  $\tan \beta$ and $D$.
In this set $m_0 $ and $ m_{\frac{1}{2}}$ are essentially free parameters.

The glunio mass reach via the jet + $\not\!\!{E_T}$ channel at the
upgraded Tevatron has been studied by Baer {\it et al.}
\cite{3}. Adopting the conventional scenario, their results can be
classified into two generic casess: i) squarks much heavier than the
gluino ($m_0 >> m_{\frac{1}{2}}$) and ii) squarks roughly degenerate
with the gluinos ($m_0\leq m_{\frac{1}{2}}$; squarks much lighter than
the gluinos are not allowed in the conventional scenario. Let us
review the results in case i) for $m_0 \simeq$ 500 GeV $>> m_{\frac{1}{2}}$.
 It was found that only $m_{\frac{1}{2}}\leq$ 75 (100) GeV 
can be probed at the upgraded Tevatron provided the integrated
luminosity accumulates to 2 $fb^{-1}$ (25$ fb^{-1}$) \cite{3}.
Unfortunately such low values of $m_{\frac{1}{2}}$ have already been
ruled out by the direct chargino searches at LEP \cite{2} and direct
squark - gluino searches by the D0 collaboration \cite{13}.  Thus
according to the conventional scenario direct squark gluino searches
at the Tevatron in the jet + $\ET$ channel will draw a blank if the
squarks indeed happen to be very heavy. This motivates us to focus our
studies on choice i) in the nonuniversal scenario. In case ii) even
the conventional scenario predicts observable signals at the upgraded
Tevatron\cite{3} and we shall not consider it further.

It may be worthwhile to mention that recent analyses of the precision
elctroweak observables have produced additional evidence, albeit
rather mild, in favour of scenario i). In \cite{14} SUSY contributions
to several of these observables were studied.  Including the
contributions from squarks, sleptons, gauginos and higgs bosons
seperately, it was found that light squarks or sleptons (with all
other sparticless rather heavy) just allowed by the current lower
limits from direct searches, always make the fit to 22 data points (Z
width and partial widths, various assymmetries etc.) worse than that of
the SM.  On the otherhand relatively light charginos and neutralinos
with heavy sfermions ($m_0>>m_{\frac{1}{2}}$) improve the fit although
the statistical significance of the improvement is rather modest.

Similar conclusions pertaining to the squark sector were obtained in 
\cite{15}. However, it was also noted that even for comparatively 
light sbottoms and small mass of one of the stops, special values of
$\tilde{t_L}-\tilde{t_R}$ mixing can make the fit as good as that in the SM.

Increasing the number of theoretical inputs the number of free
parameters can be further reduced. Several authors have noted
that \cite{16}, if Yukawa coupling unification at $M_G$ is demanded for
the third generation within an $SO(10)$ frame work, then $\tan \beta$
becomes practically fixed, since only high values of $\tan \beta$ (in a
narrow range around 50) lead to such unification. The well known
difficulty in accomodating the radiative $SU(2)\times U(1)$ breaking
in this scenario \cite{17} with universal soft breaking masses for the
scalars, may be overcome by the nonuniversality induced by the
$SO(10)$ D-terms \cite{18}.

We, however, note that the Yukawa coupling unification in SO(10) is a
consequence of the assumption that the higgs sector is indeed
minimal. In this case a single 10 dimensional higgs multiplet is
assumed to contain both the higgs doublets required to generate the
masses of the up and down type quarks and to trigger the radiative
$SU(2)\times U(1)$ breaking.  We, therefore, do not require full
Yukawa unification for the third generation and the resulting large
value of $\tan \beta$, since this crucially depends on the choice of
the higgs sector.

From the phenomenological point of view the large $\tan \beta$ scenario
in conjuction with the LEP lower bound $\MCH1\geq$ 95 GeV necessarily
implies that $\MGLU$ is almost at the edge of or beyond the
kinematical reach of the Tevatron collider. Thus direct squark-
gluino search at the upgraded Tevatron is of little consequence in
this scenario, in particular if the squarks are much heavier than the
gluino.

Even if more general higgs multiplets are assumed, b - $\tau$ Yukawa
unification is a desirable feature of the theory. The conventional
wisdom is that this requires values of $\tan \beta$ smaller than that
in the case of full Yukawa unification. Yet the favoured values of tan
$\beta$ are still too large to make gluinos sufficiently light to be
produced copiously at Tevatron energies. Typical values required by
unification are tan $\beta >$ 30. However in the presence of neutrino
masses and , in particular, of large mixing in the lepton sector this
conclusion may require revision
\cite{19,20}. 

 In the presence of large lepton mixing, as required by the SUPERK
 data on atmospheric neutrinos \cite {9}, b - $\tau$ unification can
 be achieved for relatively low values of $\tan \beta$ which were
 previously disfavoured. In fact it has been shown in \cite{20} that
 for $2 \leq $ tan $\beta \leq 4$ and suitable fermion mass matrix
 textures at $M_G$, one can obtain large mixings in the lepton sector
 along with an acceptable CKM matrix, desired neutrino mass patterns
 and b - $\tau$ unification. From the point of view of Tevatron
 phenomenology this finding is important, since gluino masses well
 within the striking range of the Tevatron are not necessarily
 excluded by the LEP lower bound on $\MCH1$ for such low values of tan
 $\beta$. We shall, therefore, restrict ourselves to the above narrow
 range of $\tan \beta$.

The sign of the parameter $\mu$ is chosen  to be  negative  since
otherwise the gluino mass range allowed by the LEP lower bound on $\MCH1$,
turns out to be uninteresting for Tevatron phenomenology.

As has been discussed in the introduction, an attractive way of fixing
the magnitude of $\mu$ is to require radiative $SU(2)\times U(1)$
breaking at the electroweak scale.  The resulting numerical value,
however, strongly depends on the choice of the higgs mass parameter at
$M_G$. Since we wish to make our predictions largely free from the
additional assumptions on the higgs sector we shall treat $\mu$ as a
free parameter. In the context of Tevatron phenomenology this,
however, does not make much of a difference since in any case
magnitudes of $\mu$ can not be much beyond 450 GeV or so, if we
require a gluino well within the striking range of the Tevatron and
$\MCH1 >$ 95 GeV. On the lower side $\mu$ is constrained by the
requirement of a bino dominated LSP.

We shall denote the cross-section corresponding to the signal with n
leptons + jets + $\ET$ by $\sigma_n$.  First we shall consider the jet
+ $\not\!\!{E_T}$ signal ($\sigma_0$) arising from squark-gluino
production.

In this work we shall reexamine the gluino mass reach at the upgraded
Tevatron for large $m_0$ ($>>m_{\frac{1}{2}})$ in the non-universal
scenario. We find that an interesting range of $m_{\frac{1}{2}}$ beyond the
LEP-2 search limit can be probed.  This is particularly so, if a high
integrated luminosity ($\sim$ 30 $fb^{-1}$) is available. We further
study the distributions of various kinematical observables associated
with the final states using conservative kinematial cuts given in
\cite{3} and compare and contrast them with the corresponding
distribution in the conventional scenario.

The size of the signal is very sensitive to the squark, glunio masses
or alternatively with $m_0$, D, and $m_{\frac{1}{2}}$.  The dependence
on the magnitude of $\mu$ and $\tan \beta$ is relatively mild but non
trivial. This variation was not studied systematically in earlier
works \cite{8}. In this paper we shall check the sensitivity of our
conclusions with respect to $\mu$ and $\tan \beta$.

If the signals for several values of D happen to be indistinguishable, we
shall try to distinguish between them by considering the distributions of
the final state observables and the corresponding dilepton ($\sigma_2$) 
and clean trilepton signals.

\section{jet + $\ET$}

In the conventional scenario Baer {\it et al.} \cite{3} have
considered the jet + $\ET$ signals in great details using the
ISAJET-ISASUSY Monte Carlo.  They have given the kinematical cuts and
the SM background corresponding to these cuts. In our parton level
Monte Carlo we have adopted the cuts and the background estimates of
Baer {\it et al.}  Although our numerical estimates based on a simple
minded approach give approximate guide lines and should not be treated
as firm predictions, the main conclusions drawn are expected to be
valid.

Baer {\it et al.} have given the jet + $\ET$ cross-section for several
representative choices of the SUSY parameters (see fig. 3 of
ref. \cite{3}). Since we are interested in the $\MSQ>>\MGLU$ case, we
have focussed our attention on the choice $m_0$ = 800 GeV, $m_{\frac{1}{2}}$ =
120 GeV, $\tan \beta$ = 2, $A_0$ = 0 and sign of $\mu$ negative. They have
also prescribed the following set of kinematical cuts $E_T(j_1)$,
$E_T(j_2)>{E_T}^c$ and $\ET >{E_T}^c$, where ${E_T}^c$ is a variable
which should be chosen appropriately for each point of the parameter
space to optimise the signal to background ratio. $E_T(j_1)$ and
$E_T(j_2)$ are the transverse energies of the two leading jets
respectively.  The other cuts from \cite{3} are $|\eta_j|\leq$3 for
all jets and $\Delta{R} (\equiv \sqrt{\Delta{\eta}^2 + \Delta{\phi}^2}) > $
0.7. Subject to these cuts the SM background is $\sim$ 2 $pb$.  In most of
the cases studied in ref. \cite{3} higher values of ${E_T}^c$ improves
the statistical significance of the signal.

 Using these cuts our parton level calculation gives cross-sections
 which approximately agree with Baer {\it et al.} for ${E_T}^c\leq$ 50
 GeV. For example for ${E_T}^c$ = 50 GeV we find $\sigma_0\approx$ 35
 $fb$ where as Baer {\it et al.} obtain $\approx$ 25$fb$. A part of the
 discrepancy ($\sim$ 10 \%) may be attributed to the use of different
 parton density functions. Baer {\it et al.} have used CTEQ2L \cite{21} while
 we have used CTEQ4M \cite{22}.  For higher values of ${E_T}^c$
 ,however, parton level calculation grossly over estimate the cross
 section compared to ISAJET result. This is understandable because the
 reduction in $p_T$ of the parton jets due to fragmentation, final
 state radiation etc.  which soften the jet $p_T$ in general, is not
 taken into account in parton level calculations. Being conservative
 we shall use ${E_T}^c$ = 50 GeV.  For a reallistic estimate we scale
 our parton level cross-sections by a factor of 2/3.  We however note
 that our conclusions regarding the search limits are likely to
 improve to some extent by the use of harder cuts.

We next present the sparticle spectrum for D = i)0.0, ii)0.4, and iii)0.6
using equations (1) and (2). The details are given in Table 1. Our main
interest will be restricted to D = 0.6 where $\MGLU>\MDR$. However, we shall
also comment on the D = 0.4 scenario.\\
\begin{center}
\begin{tabular}{|c|c|c|c|} \hline
&D=0.0&D=0.4&D=0.6\\
\hline
$\widetilde{u}_L$&550&593&614\\
$\widetilde{u}_R$&547&591&611\\
${\bf \widetilde{d}_R}$&{\bf 548}&{\bf 390}&{\bf 281}\\
${\bf \tilde{e}_L}$&{\bf 507}&{\bf 328}&{\bf 180}\\
$\tilde{e}_R$&503&550&572\\
$\tilde{g}$&312&313&313\\
$\widetilde{\chi_1}^0$&46&46&46\\
$\widetilde{\chi_2}^0$&96&95&95\\
$\widetilde{\chi}_1^{\pm}$&96&95&95\\
\hline
\end{tabular}
\vskip .3in
{\bf Table 1}: The mass spectrum in GeV at the weak scale for different values
of D with $m_0$ = 500 GeV, $m_{\frac{1}{2}}$= 105 GeV, $\tan \beta$ = 3, 
$A_0$ = 0, $\mu$ = $-$340 GeV.
\end{center}

\noindent
Further using the radiative $SU(2)\times U(1)$ breaking we find $\mu
\approx$ -340 GeV in the universal scenario (D = 0.0). In principle
$\mu$ can be determined from radiative $SU(2)\times U(1)$ breaking in
the non universal scenario as well, if we make additional assumptions
about the higgs masses at $M_G$. We shall, however, refrain from
making such assumptions and, as has already been mentioned in the
introduction, treat $\mu$ as a free parameter. In order to study the
impact of light $\DR$ squarks on the jets + $\ET$ signal we shall
first use $\mu$ = $-$340 GeV even in the nonuniversal case.  Later we
shall comment on the sensitivity of the signal to $\mu$.

For Tevatron Run II ($\cal{L} \sim$ 2 $fb^{-1}$), where $\cal{L}$
indicates the integrated luminosity, we find that for
$m_{\frac{1}{2}}\geq$ 100 GeV, no observable signal is expected for D = 0.0
in agreement with Baer {\it et al.} For D=0.4 the conclusion remains
more or less unchanged. For D=0.6 ($\MGLU>\MDR$), however, a signal
may be seen, provided $m_{\frac{1}{2}}$ is in a narrow range just
beyond the LEP II limit. For example we find for
$m_{\frac{1}{2}}$ = 110 GeV, $\sigma_0$ = 143 $fb$ which corresponds to
$\frac{\sigma}{\surd{B}}\approx$ 5.

This may be understood from the following facts. The gluino decay channels and 
corresponding branching ratio (BR)s are given in the Table 2.\\
\begin{center}
\begin{tabular}{|c|c|c|c|} \hline
&D=0.0&D=0.4&D=0.6\\
\hline
$\tilde{g}\rightarrow\CH+jets$&0.4657&0.3681&0.0006\\
$\tilde{g}\rightarrow\LSP+jets$&0.1418&0.3311&0.0001\\
$\tilde{g}\rightarrow\N2+jets$&0.3925&0.3009&0.0005\\
$\tilde{g}\rightarrow\DR+jet$&0.0&0.0&0.998\\
\hline
\end{tabular}
\vskip .5in
{\bf Table 2}: The glunio decay channels and corresponding branching ratios
for different values of D.
\end{center}

\noindent
For D = 0 case 3-body decay of the glunio dominates because all the
squarks ($\tilde{q}_L$, $\tilde{q}_R$) are heavier than the
glunio. $\CH$ and $\N2$ decay through leptonic as well as hadronic
modes. So BR($\tilde{g}\rightarrow jets + \ET$) is somewhat
suppressed. But for D = 0.4 this BR increases. This is due to the
light $\DR$ propagator which is less suppressed. As a result
BR($\tilde{g}\rightarrow\LSP+jets$) is enhanced significantly
(0.14 $\rightarrow$ 0.33).

For D=0.6, BR ($\tilde{g}\rightarrow jets + \ET$) increases rapidly.
In this case $\MGLU > \MDR$ and so the 2-body decay of glunio dominates.
First we have the decay $\tilde{g}\rightarrow \DR$ d ($\DR\equiv\DR$, 
$\tilde{s}_R$, $\tilde{b}_R$) with BR $=$ 0.998, followed by $\DR\rightarrow
\LSP d$ with 100 \% BR as the $\LSP$ is $\tilde{B}$ dominated.

The observability of jets + $\ET$ signal may improve sgnificantly if 
higher $\cal{L}~(\sim$ 30 $fb^{-1})$    
is available at the upgraded Tevatron (see Table 3).\\
\begin{center}
\begin{tabular}{|c|c|c|c|c|c|c|c|c|} \hline
\multicolumn{3}{||c}{D=0.0}
&\multicolumn{3}{||c}{D=0.4}
&\multicolumn{3}{||c||}{D=0.6}\\ \cline{1-9}
$m_{\frac{1}{2}}$&$\sigma_0$&$\frac{S}{\surd{B}}$&$m_{\frac{1}{2}}$&$\sigma_0$&
$\frac{S}{\surd{B}}$&$m_{\frac{1}{2}}$&$\sigma_0$&$\frac{S}{\surd{B}}$\\
\hline
105&53&6&105&62&7&105&203&25\\ \hline
-&-&-&-&-&-&125&53&6\\ \hline
\hline
\end{tabular}
\vskip .5in
{\bf Table 3}: The $jets + \ET$ cross-sections (in fb) and statistical 
significances
for different values of $m_{\frac{1}{2}}$ and D with $m_0$ = 500 GeV, 
$\tan \beta$ = 3, $A_0$ = 0, $\mu$ = $-$340 GeV.
\end{center}

We find that if the chargino mass is just above the LEP lower limit 
corresponding to $\MGLU \simeq$ 312 GeV, a signal may also be expected
for D = 0.0 and 0.4. For the D = 0.0 case we obtain slightly enhanced
$\frac{S}{\sqrt{B}}$ ratio compared to ref.\cite{3} since we have used
$\tan \beta$ = 3. and $\cal{L}~(\sim$ 30 $fb^{-1})$. For heavier charginos no
signal is anticipated.

For D = 0.6, however, a range of $m_{\frac{1}{2}}$ 105 GeV $\leq
m_{\frac{1}{2}}\leq$ 125 GeV can be probed. This is the consequence of the
production of relatively light $\DR$ squarks along with the gluino and
the enhanced BR ($\tilde{g}\rightarrow jets + \ET$) for reasons
discussed above.

We next study the variation of $\sigma_0$ with $\mu$. For D = 0.6 the results
hardly changes with $\mu$. This is a consequence of the fact that in this
case the decays $\tilde{g}\rightarrow\DR\, d$ and $\DR\rightarrow\LSP\, d$
dominates  the signal. The branching ratio of the former strong decay is
insensitive  to $\mu$. The second decay has $\approx$100 \% BR as long as
the LSP is $\widetilde{B}$ dominated. 

For D = 0.0 and D = 0.4 the signal has some dependence on $\mu$ (see Table 4).
\begin{center}
\begin{tabular}{|c|c|c|} \hline
$\mu$&$\sigma_0$(D = 0.0)&$\sigma_0$(D = 0.4)\\
\hline
\hline
$-$340&79&93\\
$-$400&82&80.5\\
$-$450&81&72\\
$-$500&80&66\\
$-$600&73&63.5\\
$-$700&65.5&58\\
\hline
\end{tabular}
\vskip .5in
{\bf Table 4}: The sensitivities of the $jets + \ET$ cross-section with 
$\mu$ for different values of D. All cross-sections are in fb and $\mu$
in GeV. 
\end{center}
\noindent
However for $|\mu|>$ 450 GeV, the chargino mass violates the LEP lower
bound.  Larger values of $\mu$, therefore, require enhanced
$m_{\frac{1}{2}}$ which makes the $\MGLU$ larger and the signal at the
Tevatron is supressed below the observable limit.

We next study the variation of $\sigma_0$ with $\tan \beta$.  It is
once again found that $\sigma_0$ for D = 0.6 is not at all sensitive to
this parameter.  For D = 0.4 and D = 0.0 the signal show some sensitivity.
However, for $\tan \beta \geq$ 4, the chargino mass again violates the
LEP lower bound unless $m_{\frac{1}{2}}$ and correspondingly $\MGLU$ is
increased.

\vspace*{-1.in}
\hspace*{-1.1in}
{\hbox{
\psfig{figure=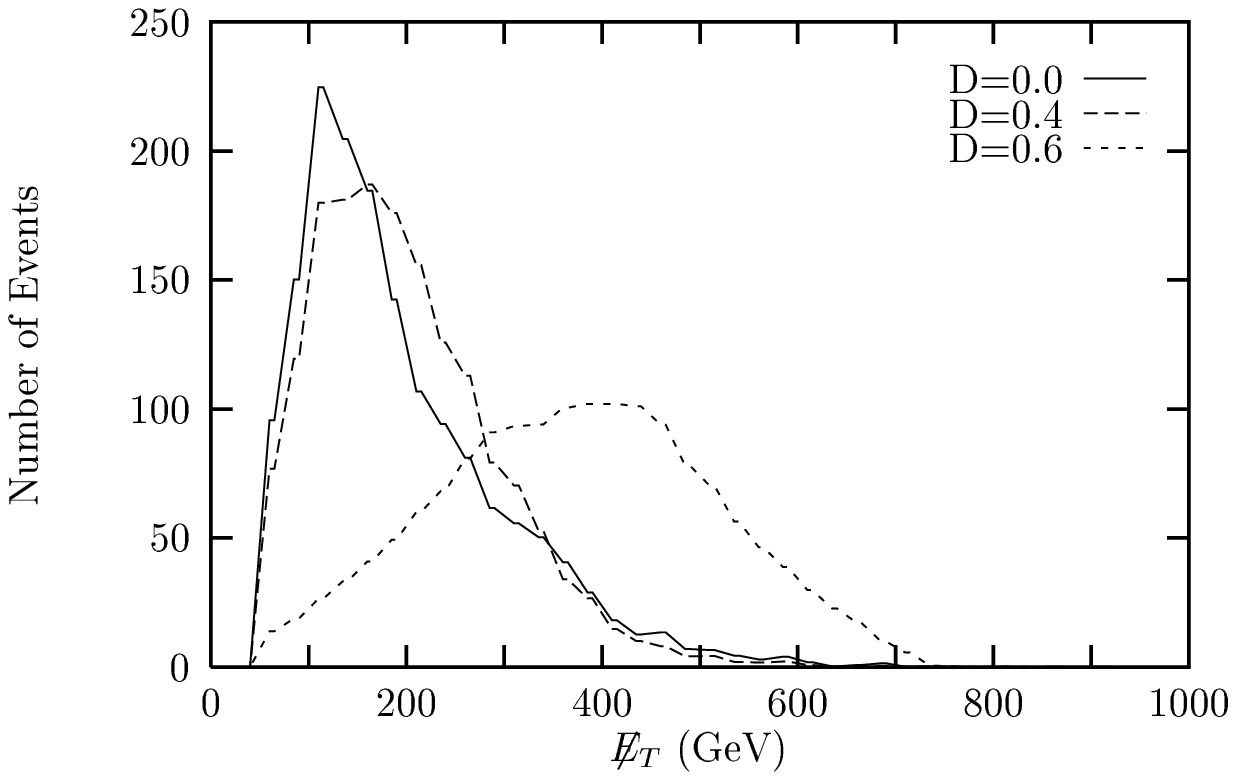,height=9in,width=7in}
}}
\vspace*{-5.in}
\begin{center}
{\bf Fig. 1}: {\it Missing $E_T$ distribution of signal for three
different values of $D$-parameter.}
\end{center}
\vspace*{.5in}
From tables 3 and 4 it follows that the magnitudes of $\sigma_0$ is
approximately the same in the following cases($m_0$ = 500, $A_0$ = 0,
$\tan \beta$ = 3):\\ 
a) $m_{\frac{1}{2}}$ = 105 GeV, D = 0.0, $\mu$ =$-$340 GeV\\ 
b) $m_{\frac{1}{2}}$ = 105 GeV, D = 0.4, $\mu$ = $-$400 GeV\\
c) $m_{\frac{1}{2}}$ = 105 GeV, D = 0.6, $\mu$ = $-$340 GeV\\ 
We now
explore the possiblity of distinguishing between these different
scenarios by using the $\ET$ distribution and the $p_T$ distributions
of the leading jet. In fig.1 we present the $\ET$ distribution with
the kinematical cuts given above. We find that already in D = 0.4 case
the distribution is considerably harder than that in the mSUGRA
scenario at least in the interval 150 GeV $\leq\ET\leq$ 300 GeV. A
bin by bin analysis of this distribution may disentangle the two
models, if a signal is seen.

\vspace*{-.5in}
\hspace*{-1.1in}
{\hbox{
\psfig{figure=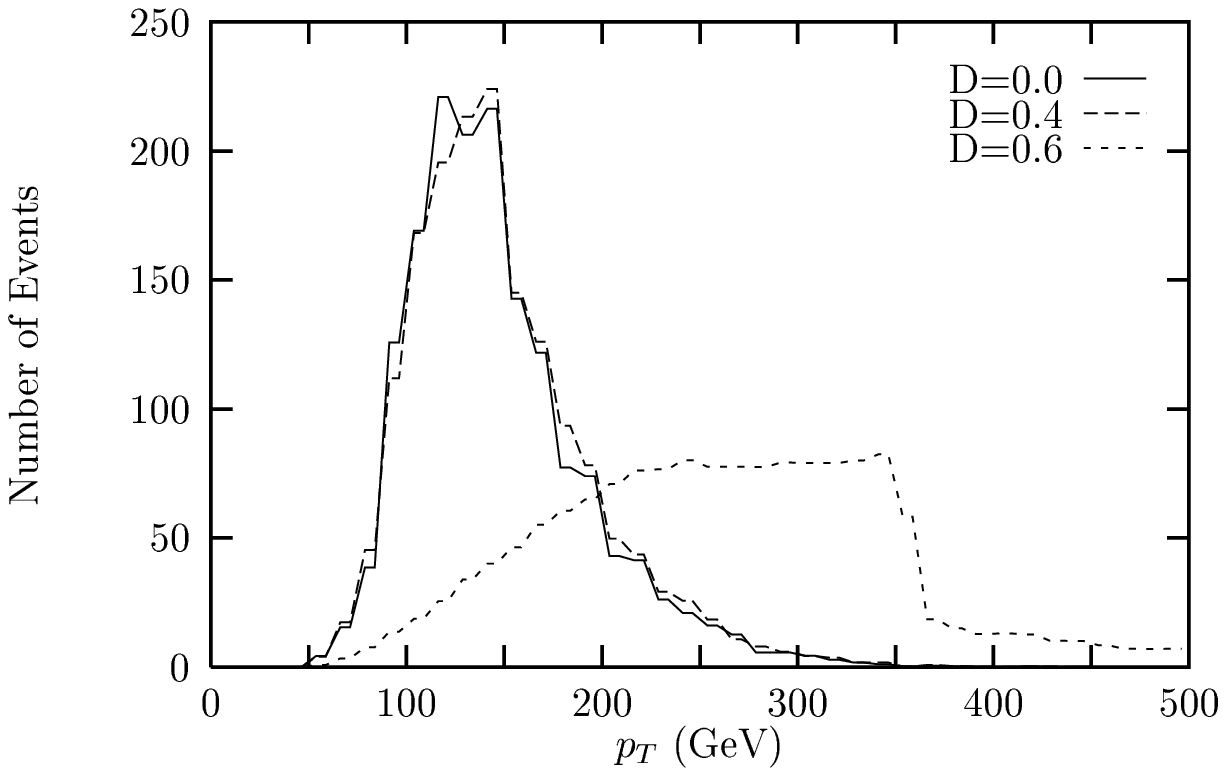,height=9in,width=7in}
}}
\vspace*{-5.in}
\begin{center}
{\bf Fig. 2}: {\it Leading-jet $P_T$ distribution of signal for three 
different values  of $D$-parameter.}
\end{center}
\vspace*{-.8in}
\hspace*{-1.1in}
{\hbox{
\psfig{figure=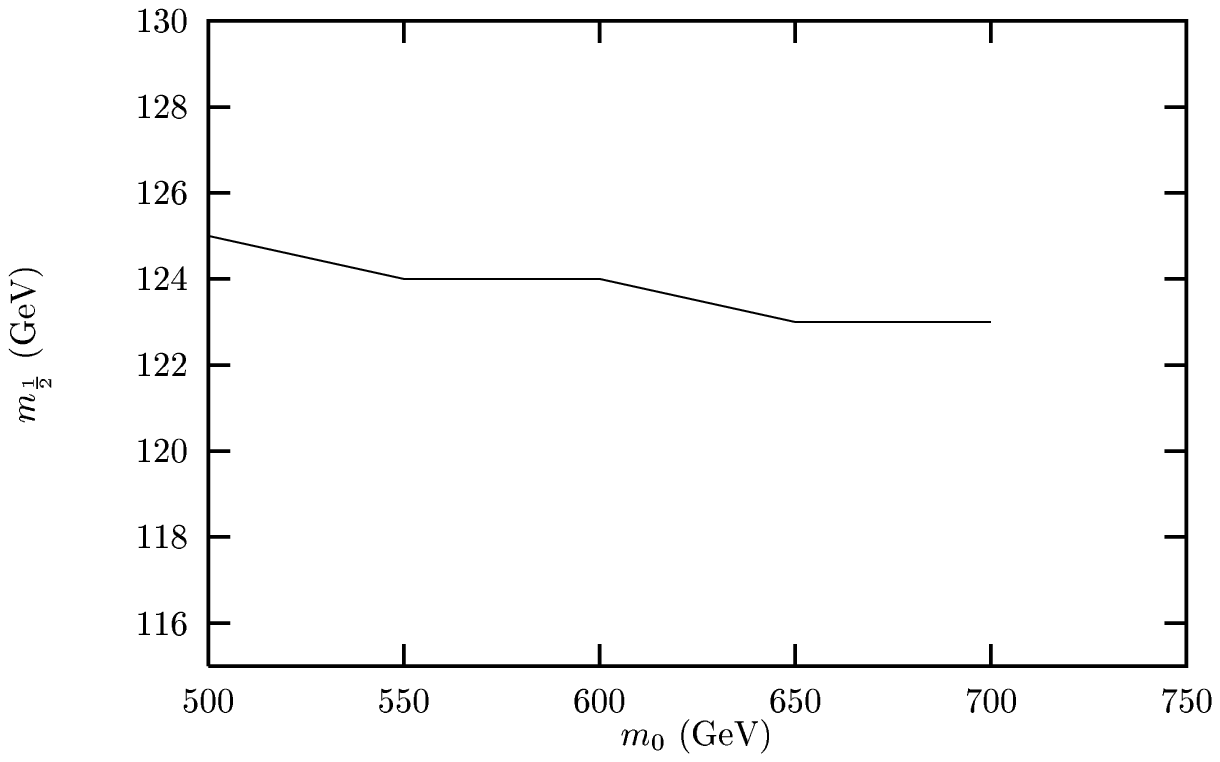,height=9in,width=7in}
}}
\vspace*{-5.in}
\begin{center}
{\bf Fig. 3}: {\it Mass reach of jets $+ \ET$ signal in $m_0 - m_{\frac{1}{2}}$
plane.}
\end{center}
 
For the D = 0.6 case the signal has a very hard $\ET$ spectrum extending
far beyond the distributions in the other two cases.

The $p_T$ spectrum of the leading jet is presented in fig.2. It is
clear that no distinction between parameter set a) and b) are possible.
A much harder $p_T$ spectrum for the leading jet, on the other hand,
can easily identify the $\MGLU > \MDR$ scenario (D = 0.6).

We next investigate the variation of $\sigma_0$ with $m_0$. As
expected $\MDR$ increases with $m_0$ and beyond a certain range
we find $\MDR>\MGLU$. Upto $m_0\sim$ 700 GeV a 5$\sigma$ signal can
be obtained for values of $m_{\frac{1}{2}}$ beyond the LEP limit
(see fig.3).

\section{OS-dileptons and clean Trileptons}

We have also studied the consequence of nonuniversality in the
opposite sign(OS) dilepton + jets + $\ET$ channel. For D=0.6,
$m_0$ = 500 GeV, $m_{\frac{1}{2}}$ = 105 GeV, $\tan \beta$ = 3, $\mu < 0$,
(i.e. $\MGLU > \MDR$) the OS-dilepton cross-section vanishes since
there is no cascade decay of the glunios. This coupled with a
relatively large $\sigma_0$ makes this model totally distinct from the
D = 0.0 case.

Baer {\it et al.} \cite{3} have studied OS dilepton signal in great
details. We have used the following cuts from \cite{3}: $E_T(j_1)$,
$E_T(j_2)\geq$ 50 GeV, $\ET \geq$ 50 GeV, $E_T(l_1)$, $E_T(l_2)\geq$ 10
GeV, $|\eta_l|<$ 2.5, and the lepton isolation criterion $R >$ 0.3.

Using the above cuts we get for $m_0$ = 800 GeV, $m_{\frac{1}{2}}$ =
120 GeV, $\tan \beta$ = 2, $\mu<$ 0, $\sigma_2$ = 1.8 $fb$ where as
Baer {\it et al.} \cite{3} have obtained $\sigma_2 \sim$ 2 $fb$. Thus
the aggrement is rather well. With harder cuts the signal/background
ratio improves. But as discussed above realistic e2stimates may not be
obtained from a parton level Monte Carlo with such strong cuts.

For $m_0$ = 500 GeV, $m_{\frac{1}{2}}$ = 105 GeV and for $\tan \beta 
\simeq$ 2  Baer {\it et al.} \cite{3}
have already analysed $\sigma_2$ in the conventional scenario. Their
conclusion was that this signal is unobservable at the upgraded
Tevatron.  However, in their analyses they asumed $|\mu|$ to be a
fixed number determined by the $SU(2)\times U(1)$ breaking condition,
which may not be realistic due reasons discussed earlier.

We have re-examined the OS signal at the parton level for D = 0.0 case
treating $\mu$ as a free parameter. We find that irrespective of the
value of $\mu$ the cross section is unobservable even with $\cal{L}$ =
30 $fb^{-1}$ in the conventional scenario.

But for $m_0$=500 GeV, $m_{\frac{1}{2}}$ = 105 GeV , $\tan \beta$ = 3 
and D = 0.4 an observable signal
may be achieved for favourable values of $\mu$ (see Table 5).
\begin{center}
\begin{tabular}{|c|c|c|c|}\hline
$\mu$&$\sigma_2$&BR($\N2\rightarrow l^+l^-\LSP$)&$\frac{S}{\surd{B}}$\\
\hline
$-$340&5.1&0.08&3\\
$-$400&7.4&0.12&4\\
$-$450&8.6&0.14&5\\
\hline
\end{tabular}
\vskip .5in
{\bf Table 5}: The sensitivities of the $OS ~dilepton + jets + \ET$ cross 
sections with $\mu$. Cross-sections are in fb and $\mu$ in GeV.
\end{center}

It is clear from the Table 5 that in the non-universal case OS
dilepton + $\ET$ may be observable for large $|\mu|$, using
$\cal{L}$ = 30 $fb^{-1}$.  This happens since $|\mu|$ increases, $\N2$
and $\LSP$ becomes more gaugino like and as BR ($\N2\rightarrow
l^+l^-\LSP$) increases.  Incidentally the OS dilepton signal may be a
convenient tool for distinguishing the D = 0.4 scenario (parameter set
in section 3) from the others which predicts unobservable dilepton
signals.  The most appropriate tool for distinguishing the three
parameter sets a, b, c presented in the previous section is
,however, the clean trilepton signal \cite{23}.

We again use the cuts of ref\cite{3}; $|\eta_l|<$ 2.5, $p_T(l_1)>$
20 GeV, $p_T(l_2)>$ 15 GeV, $p_T(l_3)>$ 10 GeV, $\ET>$ 25 GeV and
$|m(l\bar{l}) - M_Z|\geq$ 10 GeV.
In recent times it has been emphasised that the background from
$W\gamma^*/Z^*$ is the most severe one in the channel \cite{24}.
In order to take care of this background we have introduced
an additional invariant mass cut of $m_{l\bar{l}}>$ 12 GeV.
Using MADGRAPH\cite{25} we estimate the SM background to be
$\sim$ 5 $fb$ subject to the above cuts. The clean 3l cross -section
is presented in Table 6.
 
\begin{center}
\begin{tabular}{|c|c|c|c|c|}\hline
parameter &$\sigma_{\W1\N2}$&BR($\N2\rightarrow l^+l^-)$&$\sigma_{3l}$&
$\frac{S}{\surd{B}}$\\
set&in $pb$&&in $fb$&\\
\hline
a&2.294&0.019&3.0&7\\
b&2.371&0.117&16.7&41\\
c&1.187&0.152&32.3&79\\
\hline
\end{tabular}
\vskip .5in
{\bf Table 6}: The clean trilepton cross-sections in pb for different set
of parameters.
\end{center}
From Table 6 it follows that the three scenarios can be conveniently
distinguished by the clean 3l signal.

\section{Conclusions}
Within the framework of $N = 1$ SUGRA, it is quite possible that the
$\DR$ squarks are significantly lighter than all other sfermions. We
consider the case in which $\DR$ squarks have mass $\leqv$ $\MGLU$,
while all other squarks are significantly heavier than the gluino, and
outside the kinematical reach of the Tevatron. These light $\DR$
squarks have several distinctive features in comparison with the
conventional MSUGRA scenario : 1) enhancement of $j$ $+$ $\ET$ cross-section,
2) suppression of multilepton $+$ $j$ $+$ $\ET$ cross-section and the  
cascade decays of the gluinos and 3)relatively hard missing energy spectrum.

Although in view of various uncertainties in Planck and GUT scale
physics it is desirable to consider the scenario in a model independant
way, we have considered a model based on $SO(10)$ $D$-terms to generate
the mass spectra for the purpose of illustration.This model has only 
one extra parameter, namely the $D$-parameter, than the conventional 
MSUGRA model.

For $m_0$ $>>$ $m _{\frac{1}{2}}$, which  makes most of
the squarks much heavier than the gluino, it is well known that MSUGRA
does not yield an observable $j$ $+$ $\ET$ signal at the upgraded
Tevatron for gluino masses consistent with LEP or Tevatron Run-I lower
bounds \cite{2,13}.  On the other hand if $\MDR < \MGLU$, an
observable signal can be seen even with anintegrated luminosity $\sim$
$2 ~fb ^{-1}$ for $m_{\frac{1}{2}}$ just above the current lower
bound.  As the integratde luminosity accumulates to $\sim$ $30$ $fb
^{-1}$ at the upgraded Tevatron, asignificant range ( $105$ GeV $ \leq
m_{\frac{1}{2}} \leq 125$ GeV) can be probed. The variation of this
signal with $\tan \beta$ and $\mu$ is insignificant for reasons
discussed in the text \footnote{ Variation of $\mu$ is an indirect way of
taking into account the larger uncertainties in the higgs sector of
the model.}. Although we have carried out most of the calculations
for $m_o$ = $500$ GeV, similar signals can be achieved  for any $m_o$
$<$ $700$ GeV, even if $m_o$ $>>$ $m_{\frac{1}{2}}$.

We have also considered the possibilities of distinguishing between
the various scenarios from the $j + \ET$ signal at the Tevatron.
We illustrate this with three values of the $D$-parameter: 1)  $D = 0$
(conventional MSUGRA) 2) $D = 0.4$ ($\MDR > \MGLU$, but $<<
m_{\tilde q}$) and 3) $D = 0.6 $ ($\MGLU > \MDR$. We have chosen all
other parameters such that the $j + \ET$ cross-sections are comparable
in all the three cases. As discussed in th text the missing energy
spetrum in scenario 2) is already somewhat harder compared to that in
scenario 1). This difference may observable depending on the
value of the integrated luminosity. The $\ET$ spectrum in scenario 3)
is much harder compared to that in 1) and 2) and extends far beyond the 
end point of the corresponding distributions and can be easily distinguished.

Multilepton $+ j + \ET$ signals may also help to distinguish between
the three scenarios. In the scenario 3) the OS di-lepton $+ j + \ET$
is absent. In scenario 1) a slight enhancement above the SM
background is possible for all values of $\mu$, although this
enhancement is not statistically significant to qualify as a
genuine SUSY signal. In scenario 2) for suitable values of $\mu$
the OS di-lepton signal may be so large that it may be above the 
SM background in a staistically significant way.

Finally clean tri-lepton signal differ quite appreciably in the three cases.
It is the largest in case 3 while in case 2 it is still muh larger than
that in the conventional scenario.

\end{document}